\documentclass[preprint]{aastex}

\usepackage{amsmath}

\newcommand{\mean}[1]{\langle #1 \rangle}
\newcommand{\abs}[1]{\left\vert#1\right\vert}
\newcommand{\di}{\partial}
\newcommand{\divergenz}{\boldsymbol{\nabla}\cdot}
\newcommand{\unit}[1]{~\mathrm{#1}}
\newcommand{\about}{\sim\!}

\shorttitle{Grain opacity in protoplanetary atmospheres}
\shortauthors{Movshovitz and Podolak}

\begin{document}

\title{The opacity of grains in protoplanetary atmospheres}
\author{N.~Movshovitz\altaffilmark{1} and M.~Podolak}
\affil{Dept.~of Geophysics \& Planetary Science, Tel Aviv University, Tel Aviv Israel 69978}
\altaffiltext{1}{\email{naormovs@post.tau.ac.il}}

\begin{abstract}
We have computed the size distribution of silicate grains in the outer radiative region of the envelope of a protoplanet evolving according to the scenario of \citet{pollack96}.  Our computation includes grain growth due to Brownian motion and overtake of smaller grains by larger ones. We also include the input of new grains due to the breakup of planetesimals in the atmosphere.  We follow the procedure of \citet{podolak03}, but have speeded it up significantly.  This allows us to test the sensitivity of the code to various parameters.  We have also made a more careful estimate of the resulting grain opacity.  We find that the grain opacity is of the order of $10^{-2}\unit{cm^2~g^{-1}}$ throughout most of the outer radiative zone as \citet{hubickyj} assumed for their low opacity case, but near the outer edge of the envelope, the opacity can increase to $\about{1}\unit{cm^2~g^{-1}}$.  We discuss the effect of this on the evolution of the models.
\end{abstract}

\keywords{planets and satellites: formation --- solar system: formation}

\section{Introduction}
The planets in our solar system were formed from material left over after the formation of the Sun, but the details of the formation process are still under investigation. It is believed that the gas giant planets of our solar system, Jupiter and Saturn, have a solid core embedded within a gaseous envelope \citep{saumon04}. The most popular theory for the formation of these planets supposes that they were formed in two stages: A solid core was first formed by accretion of planetesimals in the protoplanetary disk, and, when the core became massive enough to gravitationally attract and capture gas from the disk, the gaseous envelope was added.

Detailed numerical simulations of these formation stages have been performed for this core accretion model \citep{pollack96,hubickyj}.  These simulations show three distinct phases. During the first phase the planet still consists primarily of solid material, the gas accretion rate is small, and the planetesimal accretion rate increases rapidly until the planet's feeding zone is depleted. The planet then enters a second phase during which the rates of solid and gas accretion are both small and nearly constant. When the gas mass of the planet is about equal to its solid mass the third phase begins, characterized by runaway accretion of gas.

Simulations of this kind, using reasonable parameters, have been successful in producing planets with properties similar to those of the actual giant planets, however the formation time for a Jupiter-like planet at $5\unit{AU}$ from a Sun-like star is uncomfortably close to the estimated lifetime of the protoplanetary disk in which it is to be formed. Observations of disks around young stars give estimates of 0.1--10~{Myr} for their lifetimes \citep{haisch01}, while numerical simulations are able to produce a planet in times varying from a few to many millions of years, depending on several key parameters \citep{hubickyj}.

One of these key parameters is the opacity of the gaseous envelope during the second of the phases described above. In particular, the contribution to the opacity from aerosols is an important factor whose value is uncertain. Grain opacity is difficult to estimate, because it depends on the size distribution as well as the composition of the aerosols.  It has been demonstrated \citep{hubickyj} that arbitrarily lowering the opacity in the outer radiative zone by a factor of 50 can reduce the formation time by roughly a factor of 3. \citet{podolak03} has shown that the size distribution of grains in the envelope can be quite different from the interstellar size distribution usually assumed in opacity calculations, and that this can result in a significantly lowered opacity.  The purpose of this work is to make this conclusion more robust by repeating the calculation with greater accuracy, and by exploring further the parameter space. We attempt to identify under which circumstances it is correct to assume that the grain opacity, in the context of the core accretion model, is much lower than the interstellar value, and to suggest the value that should be used instead.

We use the same microphysics model that was described in \citep{podolak03}, but we have made the code much more efficient so that we can more readily test the sensitivity to various assumptions.  We have also done a more careful computation of the actual opacity of the computed distribution.  In section~\ref{sec:model} we review the details of the microphysical model.  In section~\ref{sec:results} we present the results of our calculations and some tests of the sensitivity of the model to the various input parameters, and in section~\ref{sec:conclusion} we give our conclusions.

\section{\label{sec:model}Model}
\subsection{Atmospheric Model}
The model envelopes considered are taken from the work of \citet{hubickyj}. We considered three cases representing the beginning, middle, and end of phase 2 in their evolutionary scheme.  Phase 2 is by far the longest part of the evolution, and a lower opacity in this epoch will considerably shorten the evolution time.  The envelope models were kindly supplied by Dr.~O.~Hubickyj.  In order to shorten the required computation time, we considered only the outer radiative zone of the envelope.  We refer to this region as the ``atmosphere'', and it is in this region where the effect of the opacity will be most important.

\begin{figure}[p]
\includegraphics[width=\textwidth,keepaspectratio,clip]{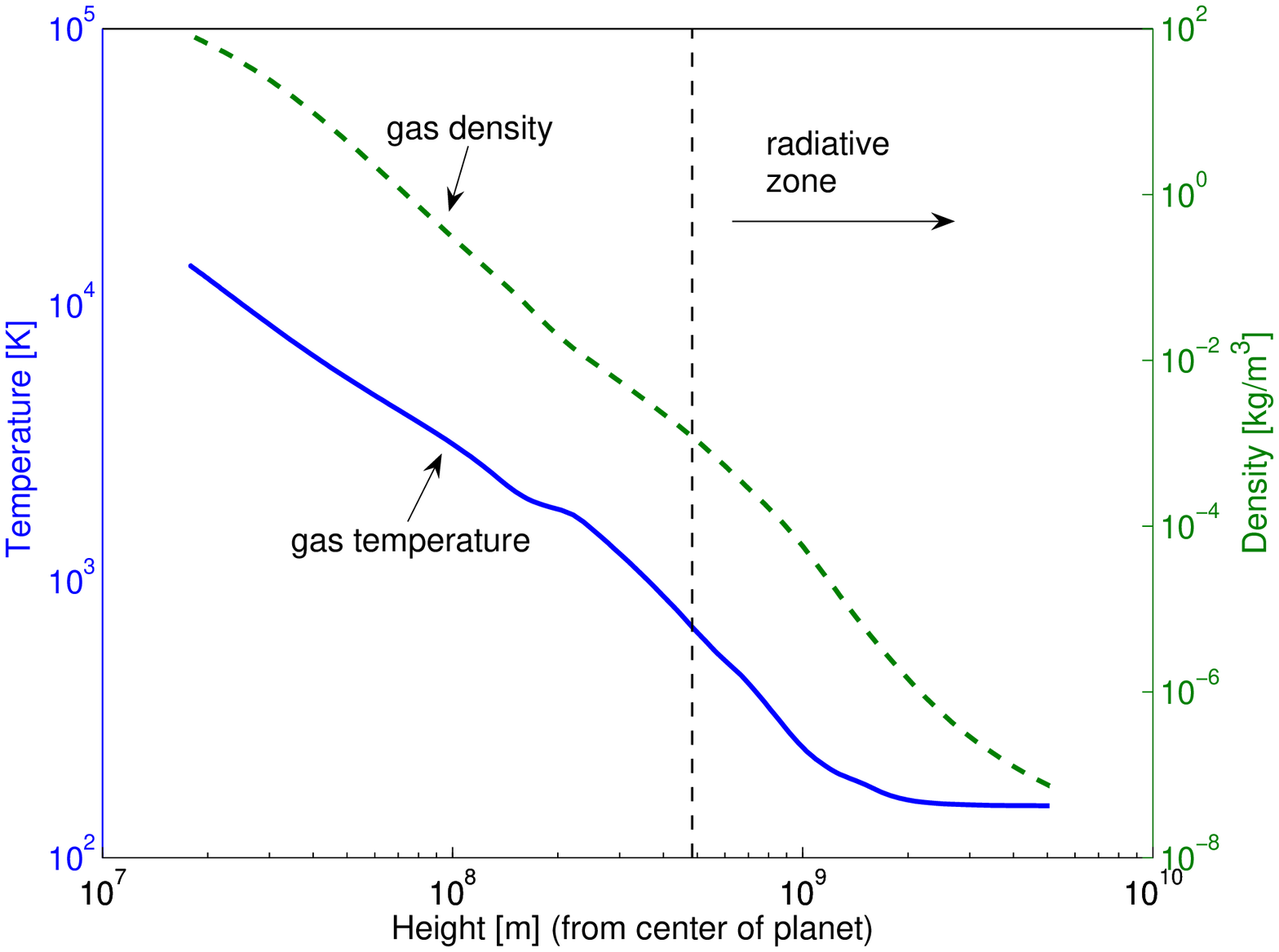}
\caption[Representative protoplanetary atmosphere]{\label{fig:atmos} A representative protoplanetary atmosphere, around the middle of phase 2: gas temperature and density as a function of height from the center of the planet.}
\end{figure}

The atmosphere and its properties change on a timescale long compared to the time it takes for the grain size distribution to reach a steady state.  As a result, we have taken the background gas properties (temperature, pressure, etc.) to be constant for each particular run.  Since we are interested in the radiative part of the envelope, we have neglected convection. A representative atmosphere model is shown in fig.~\ref{fig:atmos}. This atmosphere corresponds to a time of $0.8\unit{Myr}$, or roughly the middle of phase 2. At this time the protoplanet has a core of $12.61\unit{M_\earth}$, and an envelope of $2.73\unit{M_\earth}$.

\placefigure{fig:atmos}

\subsection{Grain Sedimentation}
We assume that the grains are solid spheres composed of some non-volatile material, so that sublimation of the grains is not considered.  The upper radiative region of the envelopes we are considering is always below a temperature of $T\approx700\unit{K}$, so that silicate grains will be stable against sublimation.  Grains composed of ice or complex organics (CHON) are not considered in the present work.  We discuss the consequences of this in the conclusion.

The sedimentation speed of a solid grain is given by 
\begin{equation}\label{eq:vsed}
v_{sed}=\frac{2\pi{a}^2\rho\,\psi{g}}{9\,\eta},
\end{equation}
where $a$ is the grain radius, $\rho$ is the grain bulk density, and $\eta$ is the gas viscosity given by $\eta=8.6\times{10}^{-7}\sqrt{T}\unit{kg~m^{-1}~s^{-1}}$ \citep{podolak88}.  The function $\psi$ depends on the Knudsen number $K\mathrm{n}$, which is the ratio of the mean free path of a molecule to the grain radius, and is given by
\begin{equation}
\psi=1+K\mathrm{n}\left[A+Be^{-C/Kn}\right].
\end{equation}
For raindrops in the Earth's atmosphere, the constants $A=1.249$, $B=0.42$, and $C=0.87$ reproduce experiments well \citep{kasten68}, and we use these values.

In reality the envelope is not static, but is constantly contracting, and we include a correction to the sedimentation velocity that is meant to approximate this.   Since there is a constant accretion of gas at the top of the envelope, we compute an effective flow speed at each level of the envelope that will provide a mass flux equal to the accreted mass flux.  We further assume that the temperature and pressure grid as a function of height remains constant for the duration of the computation and that the additional gas flows with respect to this grid. The sedimentation velocity, $v_{sed}$, is actually the speed relative to the gas flow, so the gas flow speed must be added to $v_{sed}$ to give the speed of the grain with respect to the background temperature -- pressure grid.  This additional velocity term, the same for all grain sizes, is usually negligible for grains larger than about $500\unit{\mu{m}}$ but can be comparable to $v_{sed}$ for very small grains. 

\subsection{Grain Growth}
We assume that each grain is made up of a finite number of monomers of a given size, and is spherical.  Although the grains are likely to have a fractal structure \citep{weidenschilling93} which may affect their mechanical and optical properties, we have assumed that the fractal dimension of the grains is 3 (i.e. that there are no voids).  We divide the grains among a number of size bins of radius $a_k$ such that
\begin{equation}\label{eq:sizebins}
a_k=a_02^{k/\alpha}.
\end{equation}
We have found that $\alpha=3$ gives a good compromise between numerical accuracy and computational speed.

Allowing a finite number of bins means setting an artificial limit on grain size. We have determined the number of bins necessary by running the model, without a source term, to determine the time required for $90\%$ of the mass of solids to sediment out of the atmosphere.  As we increase the number of bins and allow growth to larger particles, this ``emptying time'' generally  becomes shorter, but ultimately converges to some value that does not change when more bins are added. It is then plausible to assume that, even with an active source of small grains, adding more bins will not change the size distribution as grains will not have time enough to grow into the larger sizes. This would be true unless the source term is too large to be balanced by sedimentation, a condition that can be checked easily.

The relative probability of grains colliding as a result of their random (Brownian) motions is given by \citep[Ch. VII]{fuchs64}
\begin{equation}\label{eq:P1}
P_1(a_1,a_2)=8\pi\mean{a}\mean{D}\left[\frac{\mean{a}}{\mean{a}+\mean{\delta}\!/\,2}+\frac{4\mean{D}}{\mean{v_{T}}\mean{a}}\right]^{-1}\!\!\!.
\end{equation}
In Eq.~\eqref{eq:P1} $\mean{a}=\frac{1}{2}(a_1+a_2)$ is the average grain radius and $\mean{D}=\frac{1}{2}(D_1+D_2)$ is the average diffusion coefficient, where the coefficient for one particle is given by
\begin{equation}\label{eq:difuse}
D_i=\frac{3kT}{4\pi{a}_i^2n_{g}m_{g}v_{T}},
\end{equation}
$n_{g}$ being the number density of the ambient gas, $m_{g}$ the mass of a gas molecule, and $v_{T}$ its thermal velocity. The average thermal velocity of two grains with velocities $v_{T_1}$ and $v_{T_2}$ is given by $\mean{v_{T}}=(v_{T_1}^2+v_{T_2}^2)^{\!1/2}$. The correction factor $\delta$ takes into account the fact that a grain much larger than a gas molecule does not change its course immediately after one collision with a molecule. If $l_{B}=8D/\pi{v}_{T}$ is the distance traveled by a grain before its direction is significantly changed then
\begin{equation}\label{eq:delta}
\delta=\frac{\sqrt{2}}{6a\,l_{B}}\left[(2a+l_{B})^3-(4a^2+l_{B}^2)^{3/2}\right]-2a,
\end{equation}
and $\mean{\delta}=(\delta_1^2+\delta_2^2)^{\!1/2}$.

If there are $n_i$ particles of radius $a_i$ in a unit volume and $n_j$ particles of radius $a_j$, and if the probability of collision between any pair is independent of all other collisions, then the number $N_1(a_i,a_j)$ of collisions per second between an $a_i$ particle and an $a_j$ particle is the probability $P_1(a_i,a_j)$ times the number of different pairs, or
\begin{equation}\label{eq:N1a}
N_1(a_i,a_j)=P_1(a_i,a_j)n_in_j.
\end{equation}
The same is true for collisions between two particles of the same size except that the number of pairs is now $n_i^2/2$,
\begin{equation}\label{eq:N1b}
N_1(a_i,a_i)=\frac{1}{2}P_1(a_i,a_i)n_i^2.
\end{equation}

A larger grain can overtake a smaller one because of its higher sedimentation speed. The relative probability that a single grain of radius $a_i$ will overtake and collide with grains of radius $a_j$ per unit time is
\begin{equation}\label{eq:P2}
P_2(a_i,a_j)=4\pi\mean{a}^2\abs{v_{sed}(a_i)-v_{sed}(a_j)},
\end{equation}
and the number of such collisions per unit volume per unit time will be
\begin{equation}\label{eq:N2}
N_2(a_i,a_j)=P_2(a_i,a_j)n_in_j.
\end{equation}

Since $P_2(a_i,a_i)=0$ we can set $P=P_1+P_2$ and
\begin{equation}\label{eq:N}
N(a_i,a_j)=P(a_i,a_j)n_in_j(1-\frac{1}{2}\,\delta_{ij})
\end{equation}
for the total number of collisions.

The probability that grains will adhere after a collision is given by a  ``sticking coefficient'', $\gamma$.  In reality, $\gamma$ depends on the relative velocity of the grains (as well as other parameters). There is experimental evidence \citep{chokshi93} that for small collision speeds and small grains, the sticking coefficient is very close to one, and for most of the following we assume that the grains stick after every collision.  We have experimented with smaller (but still constant) sticking coefficients, and will present those results below.  The \emph{coagulation kernel} is then given by
\begin{equation}\label{eq:kernel}
K_{ij}=\gamma{P}(a_i,a_j).
\end{equation}
At any time $t$, the number of collisions per unit time leading to coagulation in a unit volume in a layer at height $z$, between a grain from size bin $i$ and a grain from size bin $j$ is
\begin{equation}\label{eq:bigN}
N(a_i,a_j,z)=K_{ij}(z)n_i(z)n_j(z)(1-\frac{1}{2}\,\delta_{ij}).
\end{equation}

Because there will not always be a bin corresponding to the mass of a grain formed by the coagulation of a grain of mass $m_i$ with one of mass $m_j$, we compute an array $c_{ijk}$ which gives the probability that a collision between a grain of mass $m_i$ and a grain of mass $m_j$ will form a grain of mass $m_k$.  We compute $c_{ijk}$ using the algorithm of \citet{kovetz69}.  This algorithm ensures that mass is conserved, independent of the spacing of the bins.  Further details are given in \citet{podolak03} and \citet{movshovitz}. 

A particular difficulty with this scheme is that no matter how many bins are allowed, some grains will always reach the largest bin.  These have the potential to grow larger, and thus leave the system under consideration. As a result, using the algorithm of \citet{kovetz69} will lead to mass loss from the system over time. This is not a very serious problem in terms of the resulting distribution because the number of bins was chosen in a way that ensures that most of the grains will be accommodated. But there is a simple way to correct this problem and ensure conservation of mass. After determining the desired number of bins for the distribution, say $\cal{N}$, one or two extra bins are added and used to ``recycle'' mass back into the distribution. After every time step the small amount of mass that accumulated in these extra bins is returned to the bin $\cal{N}$ and the number density of the extra bins is reset to zero. We have found that for the binning scheme that we have used (eq.~\ref{eq:sizebins}) one extra bin is enough to ensure mass conservation but more bins are needed when a finer distribution is used.

The time rate of change of the number density of grains of mass $m$ due to coagulation is given by the Smoluchowski equation
\begin{equation}\label{eq:integrodif}
\biggl[\frac{\di{n}(m,z,t)}{\di{t}}\biggr]_{coag}=\frac{1}{2}\int_{0}^{m}N(m',m-m',z,t)\,dm'-\int_{0}^{\infty}N(m,m',z,t)\,dm'.
\end{equation}
To this we add a transport term to account for sedimentation:
\begin{equation}
\biggl[\frac{\di{n}(m,z,t)}{\di{t}}\biggr]_{trans}=-\divergenz\bigl(n(m,z,t)\mathbf{v}_{sed}(m,z)\bigr),
\end{equation}
and a source term, $Q(m,z,t)$, to take into account any grains that are introduced either by the breakup of captured planetesimals, or by the accretion of gas containing grains in the outermost atmospheric layer.  The total time rate of change in the number density of grains of mass $m$ is given by
\begin{equation}\label{eq:smol}
\frac{\di{n}(m,z,t)}{\di{t}}=\biggl[\frac{\di{n}(m,z,t)}{\di{t}}\biggr]_{coag}+\biggl[\frac{\di{n}(m,z,t)}{\di{t}}\biggr]_{trans}+Q(m,z,t).
\end{equation}

The rate of planetesimal mass deposition as a function of height in the envelope was kindly provided by Dr.~O.~Hubickyj, and corresponds to the rate of planetesimal accretion used in \citet{hubickyj}.  We have converted these rates into a source term by assuming that all of the mass is deposited in the form of grains of size $a_1$ (monomers).  In addition, we assume that the gas accreted onto the planet provides an additional source of grains whose mass was usually taken to be 1\% of the accreted gas mass. For simplicity, we assumed that these accreted grains were initially of the same size as the grains from planetesimal breakup.  Once the rate of change due to the combined effects of coagulation, sedimentation, and the source term is calculated, a time step is chosen, small enough to ensure that the total change in number density at the end of the time step for every size bin and in every layer will not exceed one per cent of the number density at the beginning of the time step.

\subsection{Test of the Coagulation Algorithm}
\begin{figure}[p]
\includegraphics[width=\textwidth,keepaspectratio,clip]{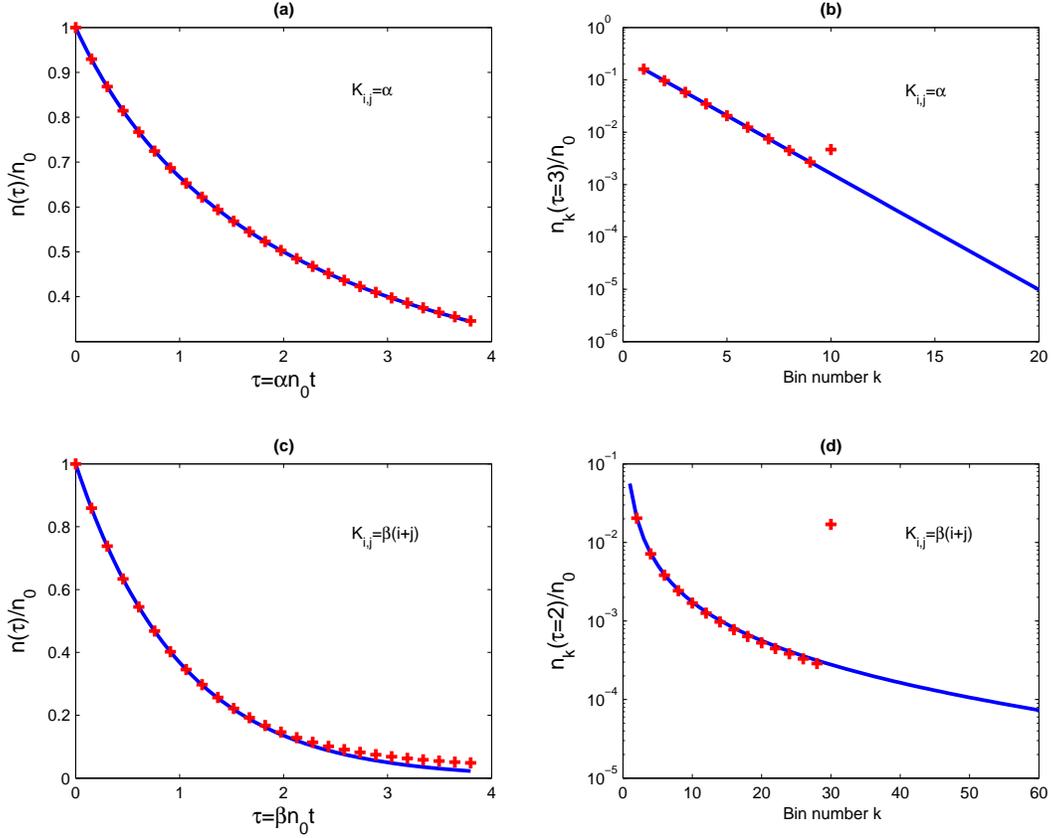}
\caption[Analytic solutions to the coagulation equation]{\label{fig:cna} Analytic solutions to the coagulation equation. In the case of a constant kernel, (a) Fraction of particles left after time $\tau$, in dimensionless time units. (b) Distribution of particles in size bins. And in the case of a linear kernel, (c) Fraction of particles after time $\tau$, and (d) Distribution in size bins. The solid lines all show the shape of the analytical function and the plus markers show the corresponding values from numerical calculation.}
\end{figure}

We have tested the coagulation algorithm by applying it to some special cases that have analytic solutions, such as the case of a constant kernel, $K_{ij}=\alpha$. With all grains initially in the smallest size bin, the total number of grains in the distribution that are left after $t$ seconds is found to be \citep{wetherill90}
\begin{equation}
n(t)=n_0(1+\frac{1}{2}\,\alpha{n}_0t)^{-1}=n_0f(t),
\end{equation}
where $n_0$ is the number of grains at $t=0$. The complete solution, the number of grains in bin $k$ for $t>0$, is
\begin{equation}
n_k(t)=n_0f(t)^2(1-f(t))^{(k-1)}.
\end{equation}
These solutions are shown in Fig.~\ref{fig:cna} along with the results of the numerical simulation.

\placefigure{fig:cna}

The numerical results agree exactly with the analytic solution. The extra grains in the last bin in the simulation (Figs.~\ref{fig:cna}b and \ref{fig:cna}d) are the result of \emph{having} a last bin, that grains can grow into but not out of. The number of grains in the last bin should be interpreted as representing all the grains of that size \emph{and larger}, or as the total number of grains in the tail of the analytic distribution.

A second case that can be solved analytically is that of a linear kernel, $K_{ij}=\beta(i+j)$ \citep{wetherill90}. The number of grains that remain after time $t$ in this case is
\begin{equation}
n(t)=n_0e^{-\beta{n}_0t}=n_0f(t),
\end{equation}
and the distribution is
\begin{equation}
n_k(t)=n_0\frac{k^{(k-1)}}{k!}f(t)(1-f(t))^{(k-1)}e^{-k(1-f(t))}.
\end{equation}
These functions are also compared with the results of simulation in Fig.~\ref{fig:cna}. Again, the numerical results agree well with the analytic solution until enough time has passed that a significant number of grains is in the tail of the distribution, above the largest bin used in the simulation.  While these tests are not applicable in the case of a real, physical kernel, they provide some measure of confidence in the numerical algorithm.

\subsection{Opacity}
The grain opacity, $\kappa_\nu$, for a particular grain size was found from the grain's extinction cross section, which was computed from Mie theory \citep{hulst57} using an approximation suggested by Dr.~J.~Cuzzi (personal communication).  If the size parameter of a grain of radius $a$ is $x=2\pi{a}/\lambda$, where $\lambda$ is the wavelength of the impinging photons, then the scattering efficiency $Q_s$ of the grain is well approximated by
\begin{equation}\label{eq:qs}
Q_s=\begin{cases}\frac{8x^4}{3}\frac{(n_r^2-1)^2}{(n_r^2+2)^2}&\text{for }x<1.3,\\2x^2(n_r-1)^2(1+(\frac{n_i}{n_r-1})^2)&\text{for }x\geq{1.3}.\end{cases}
\end{equation}
Here $n_r$ and $n_i$ are the real and imaginary refractive indices of the grain material respectively. The absorption efficiency $Q_a$ is approximated, for all grain sizes, by
\begin{equation}\label{eq:qa}
Q_a=\frac{24xn_rn_i}{(n_r^2+2)^2}.
\end{equation}
The sum of the absorption and scattering efficiencies is corrected for the degree of overall forward or backward scattering to give the extinction efficiency,
\begin{equation}
Q_e=Q_a+Q_s(1-g),
\end{equation}
where $g$, the asymmetry parameter, is approximated by
\begin{equation}
g=\begin{cases}0.2&\text{if }x<2.5\text{ and }n_i<3,\\0.8&\text{if }x>2.5\text{ and }n_i<3,\\-0.2&\text{if }x<2.5\text{ and }n_i>3,\\0.5&\text{if }x>2.5\text{ and }n_i>3.\end{cases}
\end{equation}
The effective cross-section is then
\begin{equation}\label{eq:cs}
\sigma_\nu=Q_e\pi{a}^2.
\end{equation}
This approximation gives values for the scattering and absorption efficiencies that are usually within a factor of $1.5$ of the values obtained with a full Mie theory calculation.

Equation~\eqref{eq:cs} gives the cross-section for scattering by a single grain of radiation with a given frequency. Knowing the number density of grains of this size, $n(a)$, their contribution to the opacity of the atmosphere is simply
\begin{equation}\label{eq:specific}
\kappa_\nu(a)=\sigma_\nu(a)n(a)/\rho_{gas},
\end{equation}
where we assume that the density of the gas in the atmosphere is much higher than the density of solid grains. Equation~\eqref{eq:specific} is summed over grain size to get the specific (wavelength dependent) opacity. To put our results in a form relevant to planetary evolution models, we compute the wavelength weighted Rosseland mean opacity \citep[see, e.g.,][]{clayton68}. In the remainder of the text, ``opacity'' will stand for ``Rosseland mean opacity.''

\section{\label{sec:results}Results}
\subsection{Baseline Scenario}
For the baseline model we assumed that the grains had a bulk density of $2.8\unit{g~cm^{-3}}$ and a monomer radius of $1\unit{\mu{m}}$.  The refractive indices assumed were those measured for tholins \citep{Khare84}.  We ran the model for three atmospheric profiles corresponding to the beginning, middle, and end of phase 2 in the protoplanet's evolution.

\begin{figure}[p]
\includegraphics[width=\textwidth,keepaspectratio,clip]{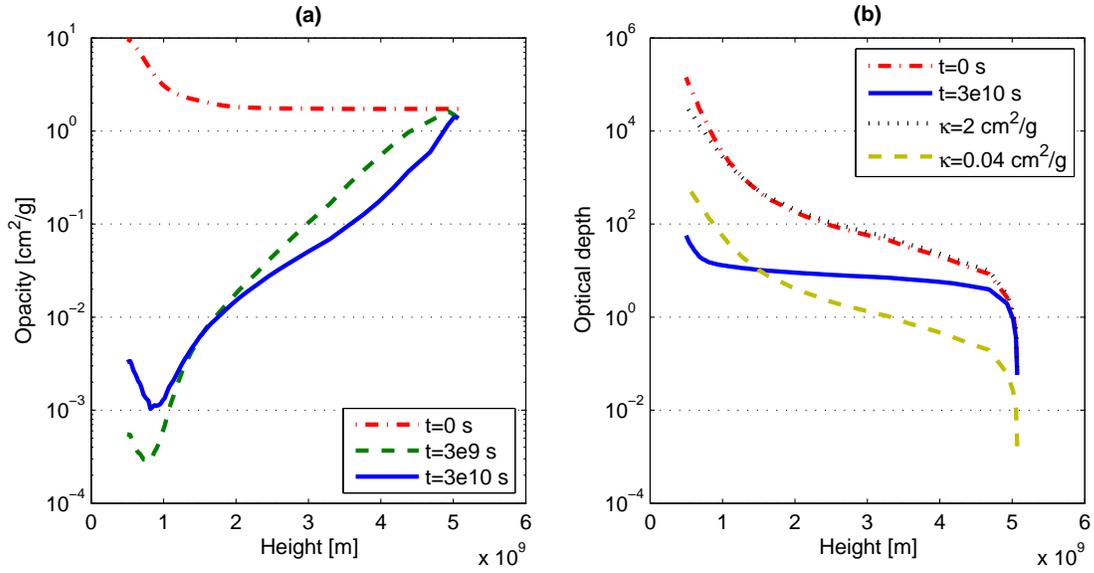}
\caption{\label{fig:5291base} Opacity (a) and optical depth (b) as a function of position in the atmosphere for three different times. Also shown in (b) are the optical depth profiles that result from taking constant \emph{High}, 2 cm$^2$ g$^{-1}$, and \emph{Low}, 0.04 cm$^2$ g$^{-1}$ opacities.  The envelope structure corresponds to the middle of phase 2.}
\end{figure}

We began the runs with 1\% of the gas mass in the form of grains. These grains were uniformly distributed with respect to the gas, and all the grains were initially monomers.  Figure~\ref{fig:5291base} shows the approach to steady state.

\placefigure{fig:5291base}

The distribution reaches a steady state after $\about3\times10^{10}\unit{s}$, or $\about1000\unit{yrs}$.  This is a small fraction of the time spent in phase 2, suggesting that the initial conditions are unimportant. Also shown in the figure are the opacity and optical depth after $10^{11}\unit{s}$, but the curve falls exactly on the curve for $3\times10^{10}\unit{s}$, and cannot be distinguished in the figure.

We see that while the opacity is quite high at first, it drops quickly when grains are allowed to grow, and the steady state opacity throughout most of the atmosphere is very much lower than $2\unit{cm^2~g^{-1}}$, the order of magnitude of interstellar grain opacity at these temperatures. Only the lower third of the atmosphere (but $\about90\%$ of the mass) is consistent with the low opacity models\footnote{\citeauthor{hubickyj}~refer to the interstellar value of grain opacity as \emph{high opacity}, and to 2\% of this value as \emph{low opacity}.} of \citeauthor{hubickyj}, but the optical depth of the radiative zone is slightly lower than it would be with their low opacity.

\begin{figure}[p]
\includegraphics[width=\textwidth,keepaspectratio,clip]{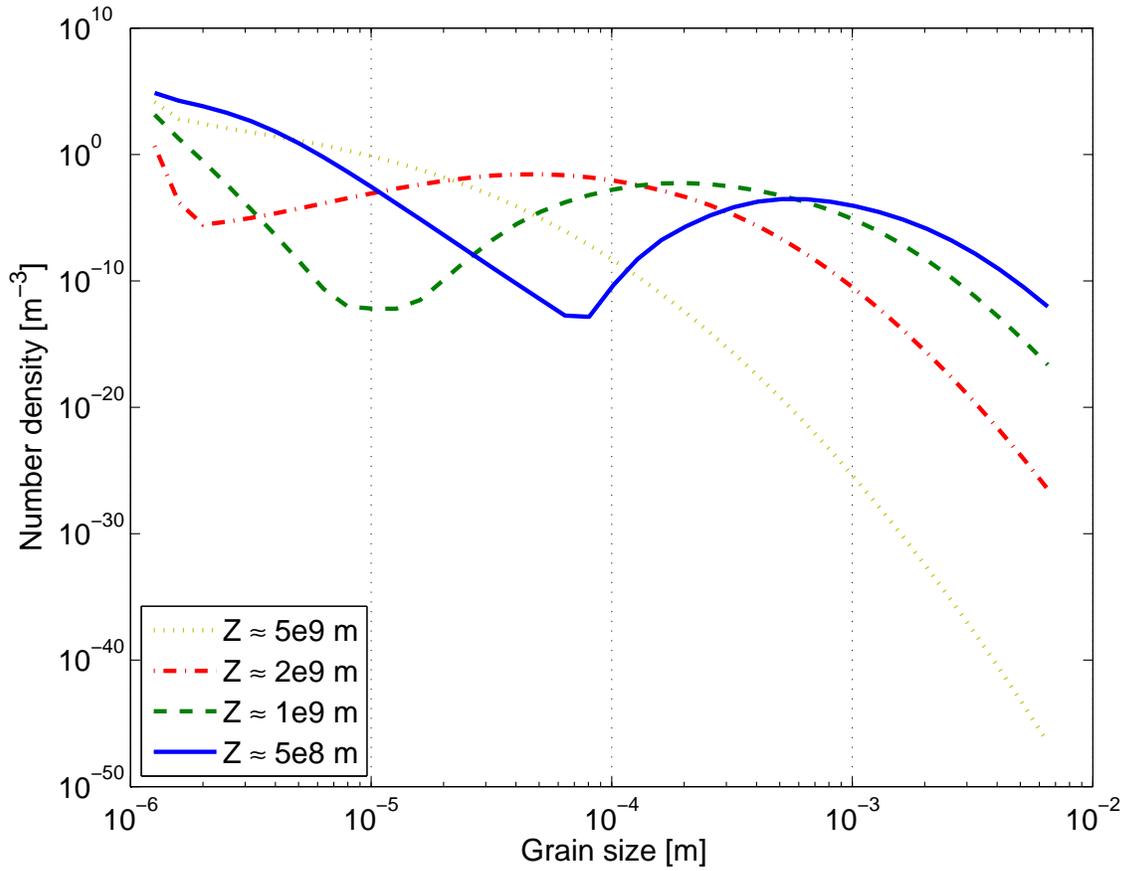}
\caption[Steady state size distribution of grains.]{\label{fig:5291nd} Steady state size distribution of grains for four levels in the atmosphere. The envelope structure corresponds to the middle of phase 2.}
\end{figure}

Figure~\ref{fig:5291nd} shows the steady state size distribution of grains in several layers of the atmosphere, at $0.8\unit{Myr}$, or roughly the middle of phase 2.

\placefigure{fig:5291nd}

Because small grains are continuously being deposited in all layers, we expect a large population of small grains everywhere. As we move deeper into the atmosphere we find more and more large grains. These are grains that had time to grow while settling from higher up. In the lower layers a significant fraction of the total mass of solid grains is in the form of grains as large as $0.1\unit{cm}$.

\begin{figure}[p]
\includegraphics[width=\textwidth,keepaspectratio,clip]{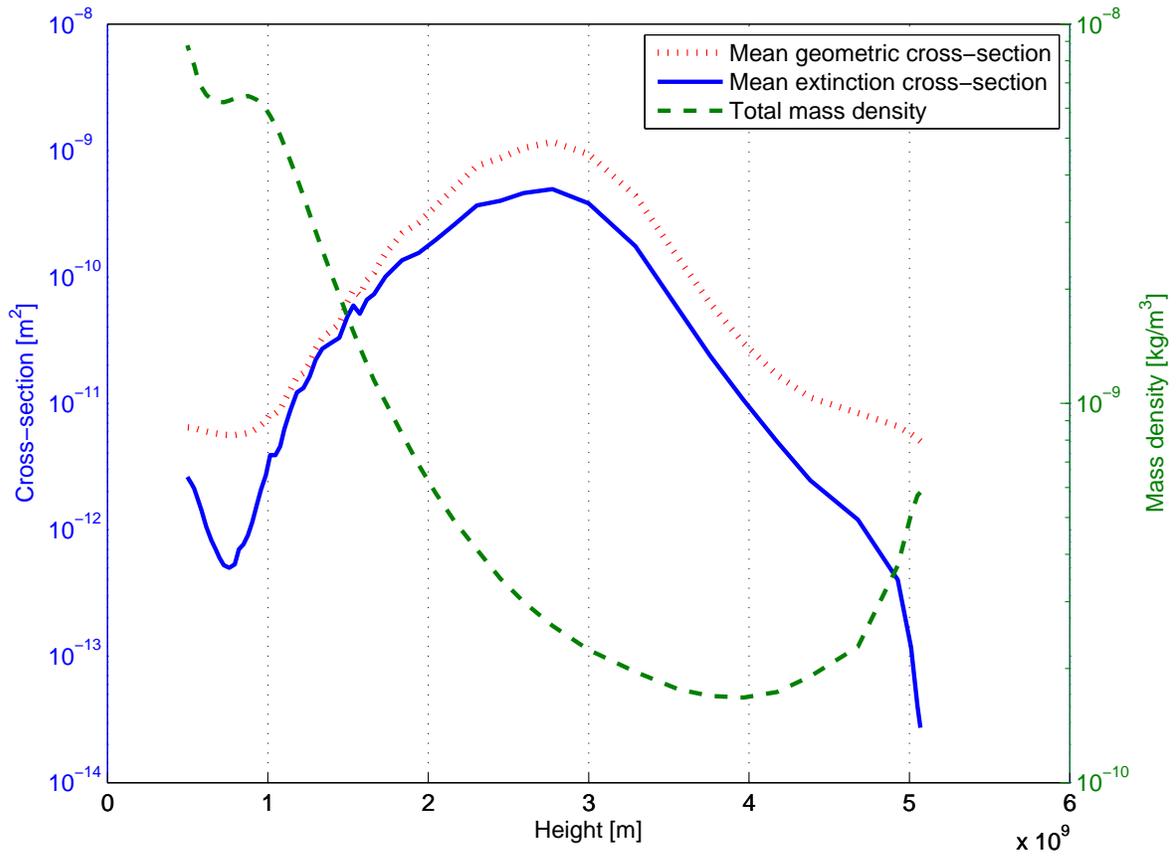}
\caption[Integrals of the steady state size distribution.]{\label{fig:5291moments}Integrals of the steady state size distribution. The extinction cross-section was calculated for the wavelength of maximum radiation at the layer's temperature.}
\end{figure}

It is instructive to look at the integrated properties of the size distribution as a function of height in the envelope,  Fig.~\ref{fig:5291moments}.

\placefigure{fig:5291moments}

Figure~\ref{fig:5291moments} shows the averages of the two quantities that determine the opacity in each layer: the mass of solid matter present as grains, and the efficiency of these grains as scatterers, shown here by the average cross-section. Note however that to correctly determine the opacity we need to know the actual shape of the distribution and not just the averages. The mean geometric cross-section, for example, is nearly identical very high and very low in the atmosphere, but the distributions (Fig.~\ref{fig:5291nd}) are very different. Figure~\ref{fig:5291moments} also shows how the grains' efficiency as scatterers, the extinction cross-section, can be very different from their simple geometric cross-section. As the temperature rises, the wavelength of the peak in the Planck spectrum decreases, and even similar size distributions can have quite different opacities.

\begin{figure}[p]
\includegraphics[width=\textwidth,keepaspectratio,clip]{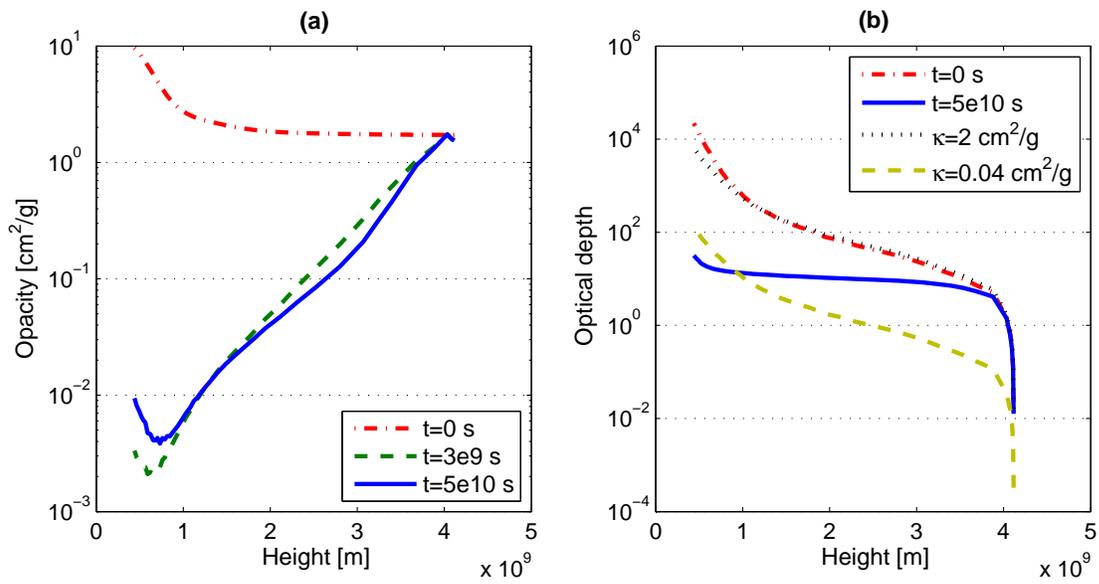}
\caption[Grain opacity and optical depth of various distributions, beginning of phase 2.]{\label{fig:2321base}Same as Fig.~\ref{fig:5291base} for a protoplanetary atmosphere at the beginning of phase 2 of core accretion.}
\end{figure}

\begin{figure}[p]
\includegraphics[width=\textwidth,keepaspectratio,clip]{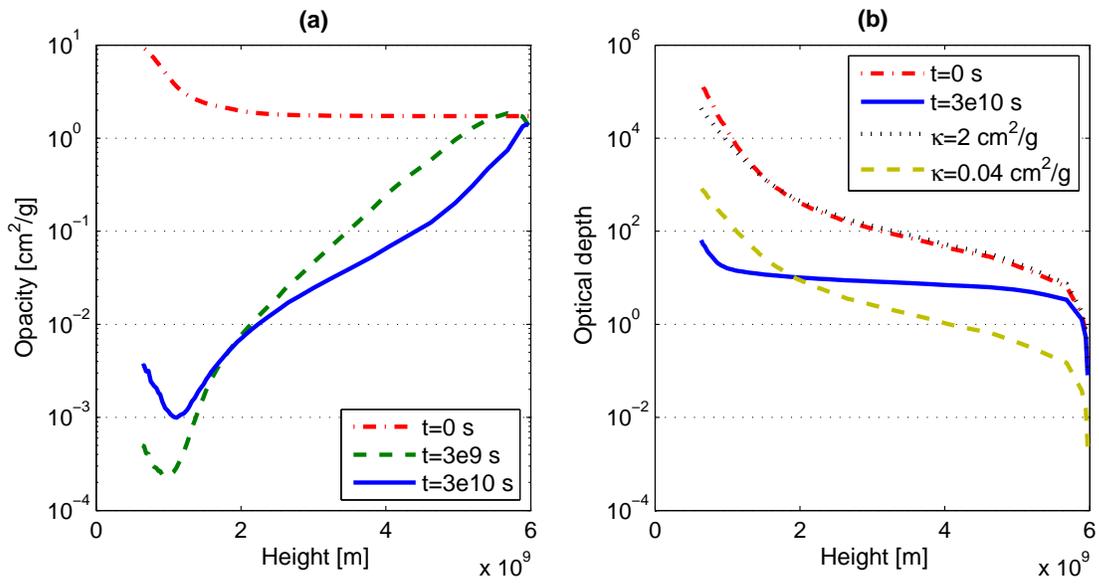}
\caption[Grain opacity and optical depth of various distributions, end of phase 2.]{\label{fig:8651base}Same as Figs.~\ref{fig:2321base} and \ref{fig:5291base} for a protoplanetary atmosphere at the end of phase 2.}
\end{figure}

Figures~\ref{fig:2321base} and~\ref{fig:8651base} show the opacities at roughly the beginning, $0.35\unit{Myr}$, and the end, $1.3\unit{Myr}$, of phase 2 respectively.

\placefigure{fig:2321base}
\placefigure{fig:8651base}

The envelope and core of the protoplanet grow more massive with time, but the structure of the atmosphere, the $(z,T,\rho)$ relation, is not radically different. The corresponding opacities show the same behavior that was observed in Fig.~\ref{fig:5291base} -- the steady state opacity is highest at the top of the atmosphere, where it is similar to the interstellar opacity, and decreases gradually bringing the lower third of the atmosphere to less than two per cent of the interstellar value.

\begin{figure}[p]
\includegraphics[width=\textwidth,keepaspectratio,clip]{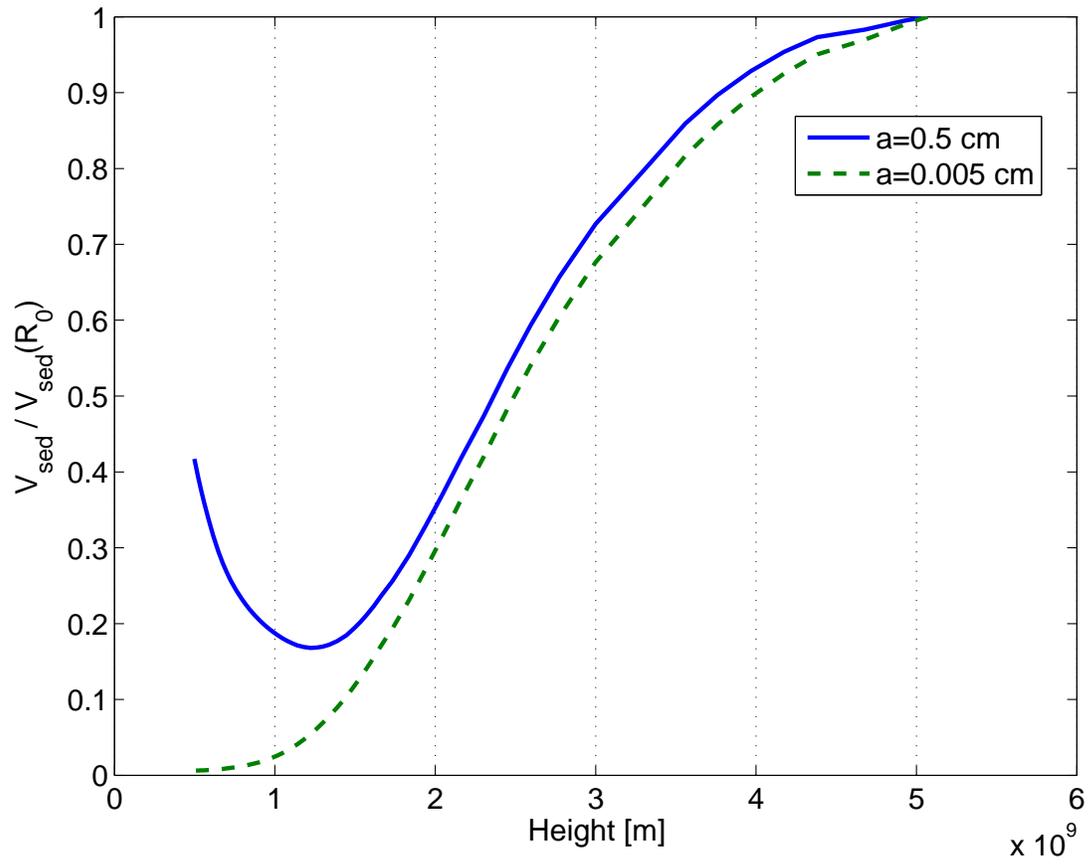}
\caption[Trends of the sedimentation velocity of grains, comparing small grains with large ones.]{\label{fig:5291vsed}Trends of the sedimentation velocity of grains, comparing small grains with large ones.}
\end{figure}

 The small increase in opacity in the very last few layers of the atmosphere is puzzling at first. We see a similar feature in Fig.~\ref{fig:5291moments}, with the increase of the mean extinction cross-section at approximately the same height. A clue for understanding this feature can be gained by looking at the behavior of the sedimentation velocity of larger grains,~Fig.~\ref{fig:5291vsed}.

\placefigure{fig:5291vsed}

In the upper atmosphere the drag force on all grains is given by the formula for  large Knudsen number and is proportional to the gas density. The gas density increases rapidly as the grains move deeper into the atmosphere.  This increases the drag force and slows the sedimentation. A short distance above the bottom of the radiative zone the gas density is high enough so that the \emph{largest} grains enter the regime of low Knudsen number. The drag force is then proportional to the viscosity, which changes only as the square root of the temperature. As a result, the drag force is almost constant, and the increasing pull of gravity causes an increase in sedimentation velocity of larger grains, but not of smaller ones. This effect creates a small shift in the size distribution, increasing somewhat the opacity.

\subsection{Monomer Size}
\begin{figure}[p]
\includegraphics[width=\textwidth,keepaspectratio,clip]{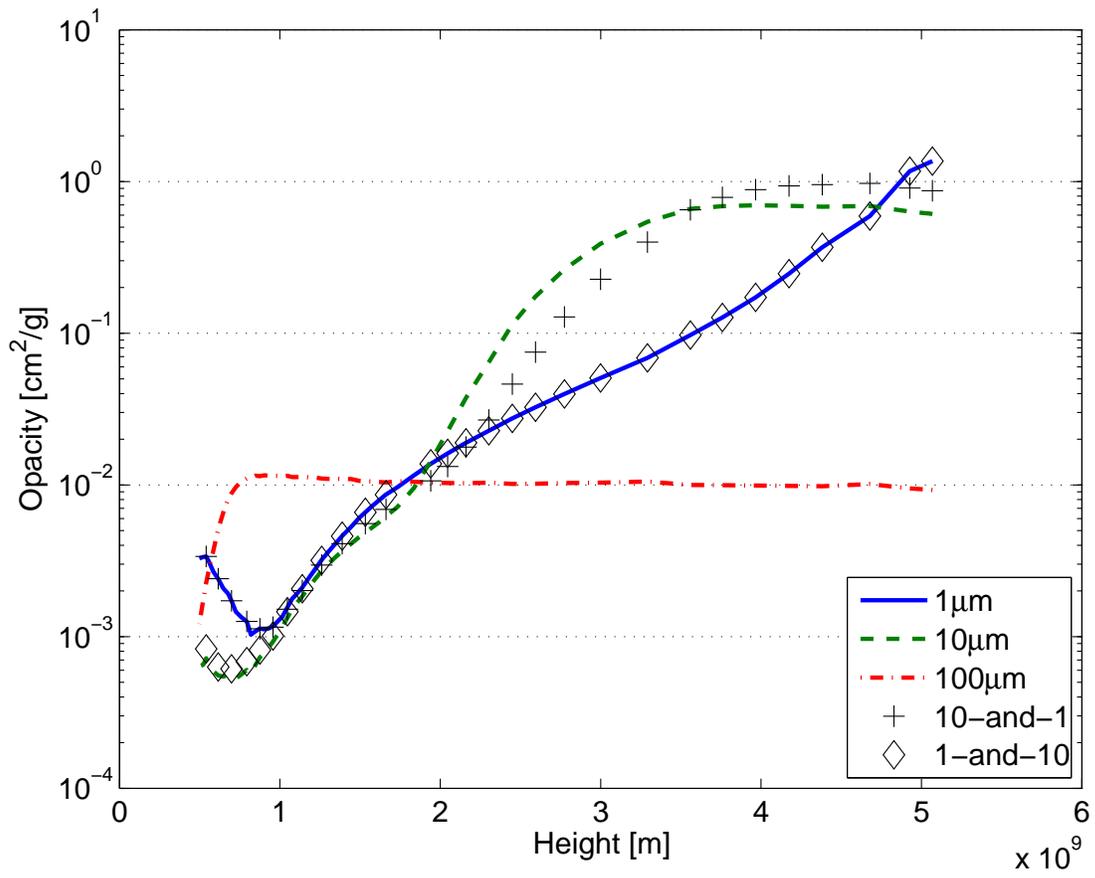}
\caption[Effect of monomer size on the steady state opacity.]{\label{fig:monos}Effect of monomer size on the steady state opacity.}
\end{figure}

The sensitivity to the choice of monomer size is shown in Fig.~\ref{fig:monos}, which shows the steady state opacity profile for the middle of phase 2, at $0.8\unit{Myr}$.  Plotted are the steady state opacity profiles calculated with  different choices of $a_0$.  In the first three cases, of $1$, $10$, and $100\unit{\mu{m}}$, the input grains, both from planetesimals and from nebular gas, are assumed to be composed of monomers only. The time to reach steady state increases somewhat when the monomers are larger, but the size of the largest grains in the steady state distribution does not seem to depend on the monomer size. If $0.1\unit{\mu{m}}$ monomers are used then the resulting opacity profile  coincides almost exactly with the $1\unit{\mu{m}}$ (baseline) case.

\placefigure{fig:monos}

The monomer size is obviously an important parameter. This is mostly due to the assumption that the constant influx of gas and planetesimals contains grains only of this size. This constant flux sets a kind of ``boundary condition'' on the steady state distribution that determines the concentration of monomer grains in every layer. The concentration of monomer grains determines, in turn, the rate of growth and thus the shape of the entire distribution. In particular, $10\unit{\mu{m}}$ monomers produce much higher opacity than both larger and smaller monomers. The $10\unit{\mu{m}}$ grains are large enough to be efficient scatters yet small enough to have a high surface area per mass ratio, while their lower number density than that of smaller grains limits the collision efficiency due to Brownian motion, the dominant collision process in the upper layers where grain populations are still homogeneous. The curve for $100\unit{\mu{m}}$ monomers, on the other hand, is almost flat over most of the atmosphere, because their large size means a low number density and collisions occur at a very slow rate. Surprisingly, the different opacity profiles seem to almost converge in the lower layers of the atmosphere, and in all cases the opacity is lower than interstellar.  In the lower third of the atmosphere the opacity is consistent with the low opacity assumption of \citeauthor{hubickyj}.

The two cases marked 1-and-10, and 10-and-1, show the result of separating grain input by planetesimals from those carried in by nebular gas. The 1-and-10 case (in diamond markers) corresponds to having $a_0=1\unit{\mu{m}}$ for the grains entering with the gas, and $a_0=10\unit{\mu{m}}$ for the grains coming from planetesimals. The 10-and-1 case (in plus signs) is just the reverse. These cases demonstrate the fact that planetesimal grains, because they are released mostly in the deep atmosphere, only influence the size distribution in the deep layers. While grains carried in by nebular gas, because they are deposited in the uppermost layer, influence the size distribution throughout most of the atmosphere.

The fact that the opacity profiles are sensitive to the choice of monomer size is perhaps a drawback of the procedure used to calculate them. In particular there is no reason that the source of new grains will contain only monomers, or even contain monomers at all, but since our model does not include processes that \emph{reduce} the grain size, like breakup or evaporation, there is no reason to consider any grains smaller than those brought in from outside the planet. In future work we hope to include these processes.

\subsection{Sticking Coefficient}
\begin{figure}[p]
\includegraphics[width=\textwidth,keepaspectratio,clip]{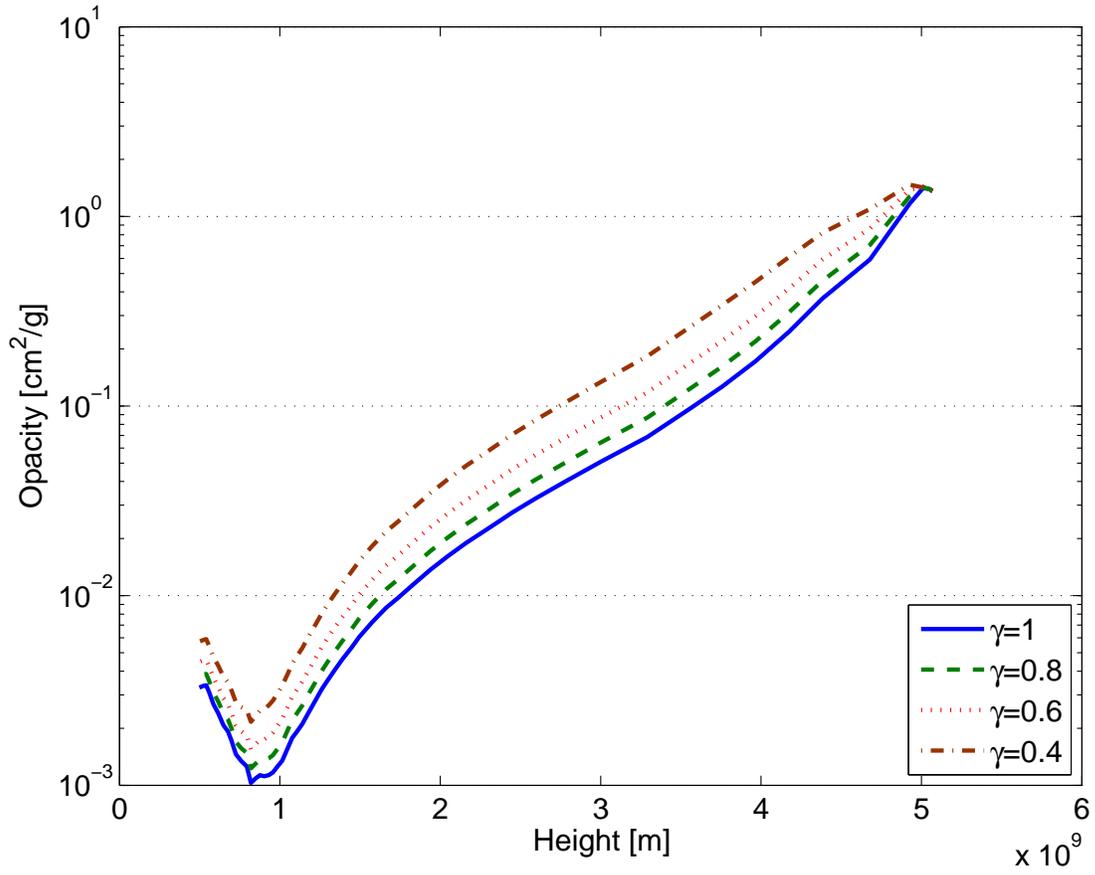}
\caption[Effect of sticking coefficient on steady state opacity.]{\label{fig:stick}Effect of changing sticking coefficient on steady state opacity.}
\end{figure}

The sensitivity to the sticking coefficient, $\gamma$, is shown in Fig.~\ref{fig:stick}.  It turns out that the resulting opacity is mildly sensitive to changes in this parameter. The opacity profiles in Fig.~\ref{fig:stick} show the same behavior, with approximately the same values, for a range of sticking coefficients from $1$ to $0.1$.  As can be seen, the opacity and optical depth increase somewhat as the sticking coefficient decreases by an order of magnitude, but the overall values of the opacity are still low.  The time to reach steady state when $\gamma=0.1$ is $\about 3500$ yrs, as opposed to $\about 1000$ yrs for $\gamma=1$.  This is still much lower than the time scale for significant changes in the envelope structure.

\placefigure{fig:stick}

\subsection{Planetesimal Size}
During most of phase 2 of core accretion the protoplanet gains on the order of $10^{-6}\unit{M_{\oplus}~yr^{-1}}$ of high-Z material from accreting planetesimals. But most of this mass is added to the planet's atmosphere \emph{below} the radiative zone, or possibly added directly to the core. In their model of planet formation, \citeauthor{hubickyj} calculate first the rate of solid planetesimal accretion by the protoplanet, then calculate the interaction between a single planetesimal and the planet's atmosphere to determine how much mass and energy are deposited at different heights. They assume, for the purpose of this calculation, that the planetesimals all have a radius of 100 km. In general, smaller planetesimals are expected to deposit more of their mass higher up in the atmosphere while larger planetesimals deposit their mass deeper in the atmosphere or even survive to reach the core nearly intact.

\begin{figure}[p]
\includegraphics[width=\textwidth,keepaspectratio,clip]{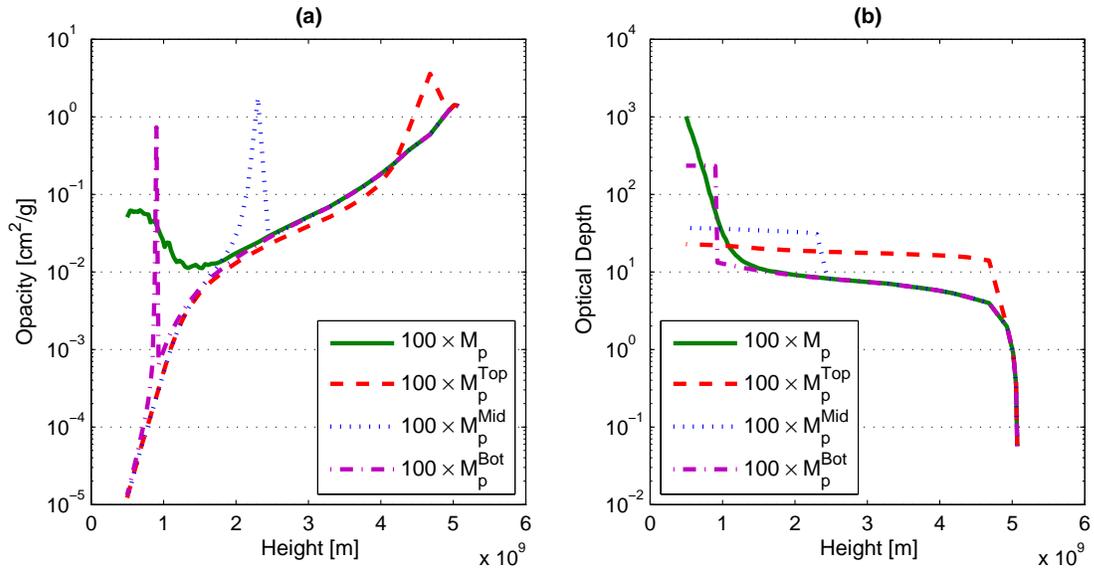}
\caption[Increased input from planetesimals.]{\label{fig:tesims} Increased input from planetesimals, accounting approximately to the possibility of smaller planetesimals. The line marked $100\times{M}_p$ corresponds to increasing the planetesimal dust source by a factor of hundred, throughout the radiative zone, maintaining the mass deposition profile. The other lines correspond to placing the increased source of grains all in one layer, at approximately the top, the bottom, and the middle of the radiative zone.}
\end{figure}

Rather than repeat these lengthy calculations with planetesimals of different sizes, we have  tested the effect of varying the assumed planetesimal size by varying directly the mass deposition profiles. Fig.~\ref{fig:tesims} shows the results.  The heavy solid curve shows the effect of increasing the value of the source term by a factor of 100 throughout the atmosphere.  The increased input of monomers means that there will be many more small scatterers, and the opacity will increase.  The higher number density of monomers also leads to quicker growth and settling, so that the opacity goes up by no more than an order of magnitude in the deepest layers of the atmosphere for a two order of magnitude increase in the source term.

Also shown in the figure are the results for the same case, but with all of the material being deposited in some particular layer in the atmosphere.  This might be the case if the planetesimals were particularly small, and broke up high in the atmosphere.  As can be seen from the figure, there is a peak in the opacity at the level where the material is put into the atmosphere, but the high concentration of grains  at that level leads to growth and sedimentation, and the resulting opacity deep in the atmosphere actually decreases.

\placefigure{fig:tesims}

\subsection{Other Parameters}
\begin{figure}[p]
\includegraphics[width=\textwidth,keepaspectratio,clip]{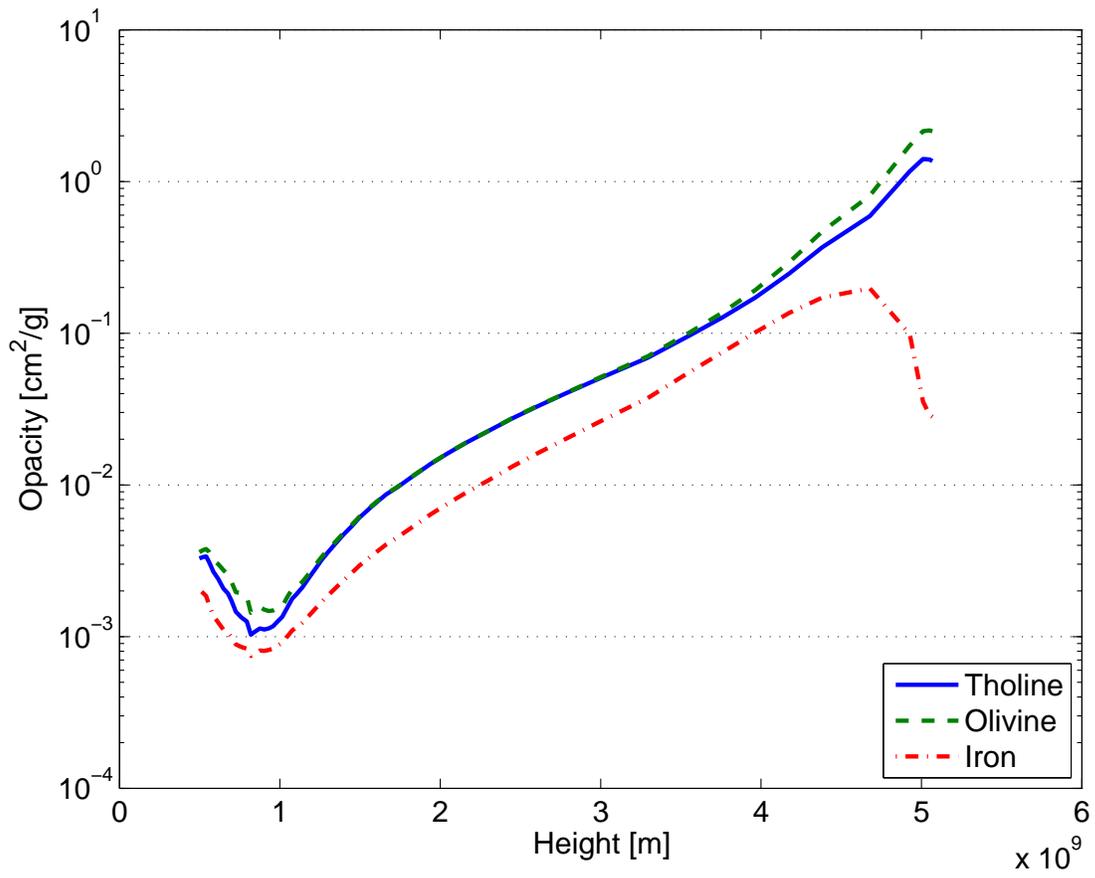}
\caption[Comparison of the opacities of tholin, olivine, and iron grains.]{\label{fig:oliv}Comparison of the opacities tholin, olivine, and iron grains.}
\end{figure}

In the baseline scenario the grain opacity is calculated using a table of wavelength dependent refractive indices measured for tholins \citep{Khare84}. Figure~\ref{fig:oliv} compares this opacity profile with one calculated with a refractive index table measured for olivine \citep{dorschner95}. As can be seen, the opacity is not sensitive to the exact values chosen for the refractive index. Also shown in Fig.~\ref{fig:oliv} is the opacity profile for grains made of iron. In our model this is accomplished assuming a bulk density of $7.87\unit{g/cm^3}$ for the grains, as well as using a refractive index table appropriate for iron \citep{weaver79}.

\placefigure{fig:oliv}

\begin{figure}[p]
\includegraphics[width=\textwidth,keepaspectratio,clip]{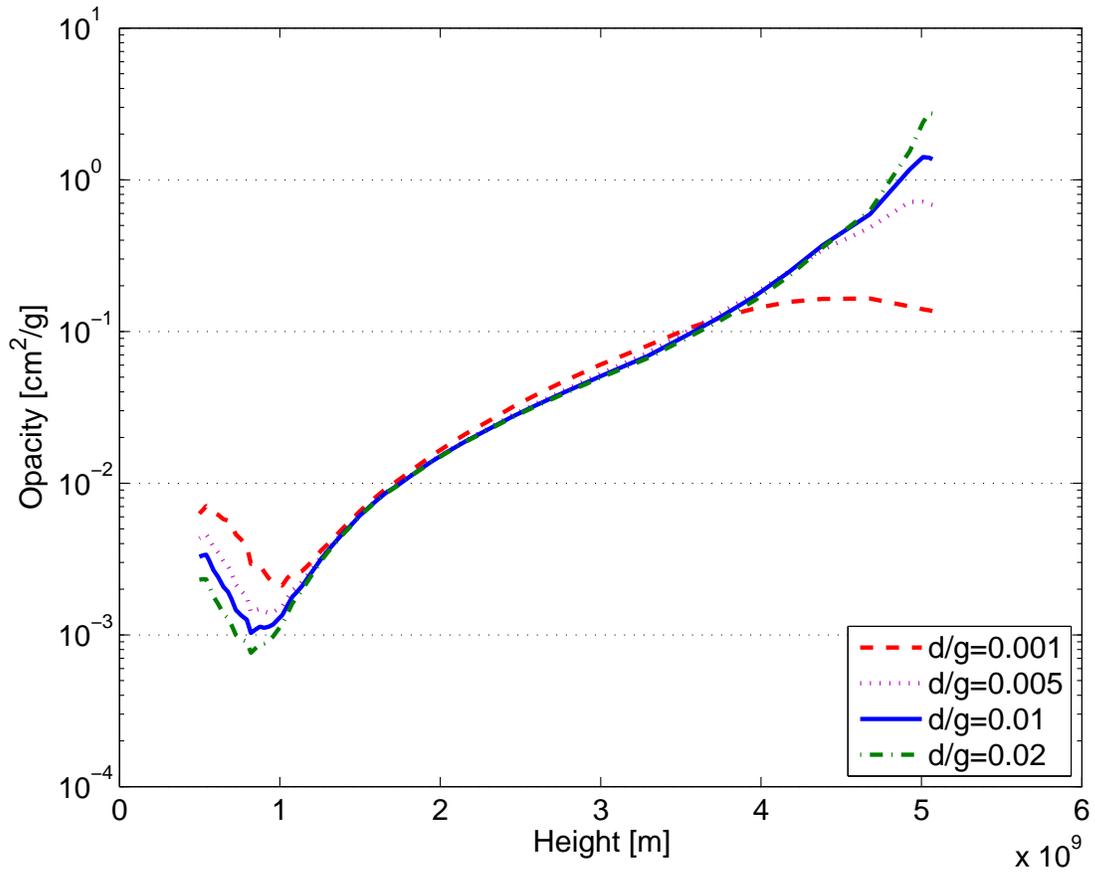}
\caption[Comparison of opacity profiles obtained with several ratios of solid matter in nebular gas.]{\label{fig:d2g}Comparison of opacity profiles obtained with several ratios of solid matter in nebular gas.}
\end{figure}

The amount of solid matter that is contained in the nebular gas that is accreted by the planet should be close to 1\% of the gas mass. Figure~\ref{fig:d2g} compares the opacity profiles obtained when this ratio is varied.  Although, for the most part, the opacity is insensitive to factor of two changes in this value, there is a fairly large opacity \emph{increase} in the bottom of the atmosphere when the grain content of the accreted gas is \emph{reduced} by an order of magnitude. This is again related to the efficiency of the grain growth process, which is dependent, in a nontrivial way, on the interdependence of the growth rate and the sedimentation rate on the concentration of small grains, dictated by the source term. In this way, while the mass of grains in the atmosphere may be less, the opacity in some layers may be higher.

\placefigure{fig:d2g}

\section{\label{sec:conclusion}Conclusions}
Before any conclusions can be drawn from the above results about the effect of grain growth on planet formation it is important to note again one important assumption of the calculations made here, the assumption of a static atmosphere. As was previously mentioned, the radius-density-temperature relation of the atmosphere remains constant throughout the calculation. This assumption was justified because the time to reach a steady state of the grain size distribution is much shorter than the time scale of the atmosphere's evolution. We must remember however that the structure of the atmosphere which is used in the present calculation was determined \emph{without} knowledge of the grain size distribution, and with previously chosen grain opacity. The grain opacity itself will, of course, influence the evolution of the planet's atmosphere. In this sense, the calculations carried out here are not completely self consistent. It is possible that using the opacity profiles shown above in a planet evolution model would result in different atmosphere structures which, if used as background for the grain growth process, would result in different opacity profiles.

With this reservation in mind, there is at least one conclusion that can still be stated with some confidence about the effects of grain growth on the atmosphere's opacity, and that is that grain growth \emph{does affect} the atmosphere's opacity. More precisely: The conditions during phase 2 of core accretion are such that growth and settling are efficient processes by which grains are removed from the radiative zone of the planet's atmosphere, reducing their contribution to the opacity. For most of the radiative zone the resulting grain opacity is significantly lower than it would be if grain growth were not accounted for. The implication for giant planet formation theory would be that the formation time of a Jupiter-like planet through the core accretion scenario may be shortened, alleviating one of the major points of concern about this model.

As for the actual value of grain opacity that should be used in planet evolution models, our simulations indicate that the opacity is not reduced by a constant factor throughout the atmosphere.  Instead, it is roughly equal to the interstellar grain value in the uppermost layers, where the grains reside for a relatively short time, and grain growth is not an important process.  Deeper in the atmosphere, where the grain microphysics has time to adjust the grain size distribution, the opacity can be reduced over the interstellar value by as much as three orders of magnitude.

In the uppermost parts of the atmosphere, where the optical depth is $\tau\leq\sim 10$ the optical depth profile is close to that computed for the high opacity case of \citet{hubickyj}, but as you descend into the deeper layers of the radiative zone, the optical depth remains well below the profile derived for the high opacity case.  Except for the deepest region of the radiative zone, the optical depth in our models is always higher than that derived for the low opacity case of \citet{hubickyj}.  It seems that we must combine a microphysics calculation of the grain distribution together with an evolutionary model of the sort presented in \citet{hubickyj} in order to determine the true effect of the grain size distribution on the contraction time of the protoplanetary envelope.

In the present work, evaporation of grains was completely neglected. If there are ice grains present, they will evaporate once the temperature exceeds $\about 200$ K. Organic material in the grains will either evaporate or pyrolyze in the hydrogen-rich atmosphere. Ice and organics are expected to make up about $2/3$ of the high-Z material by mass, so the grain abundance throughout most of the region we have considered will be only about 30\% of what we have considered for the baseline model.  As we showed above, such a reduction in the rate of grain accretion has only a small effect on the resulting distribution.  However, it must be remembered that in the upper layers, where the temperatures are low enough, grains can accumulate ice and organics.  These components will be lost in the deeper layers.  If the ice acts as a ``glue'' to hold the silicates together, then we would expect the larger grains to break up upon reaching the deeper, hotter layers.  At the very least the grains will develop voids and possibly a fractal structure.  This will lower their mean density and have a substantial effect on the size distribution.  Thus our conclusions about the opacity must be seen as tentative, subject to future work.

\acknowledgments
we would like to thank two anonymous referees for
their helpful comments.

\bibliographystyle{icarus2}
\bibliography{paper}

\end{document}